\documentclass{article}

\usepackage{arxiv}

\usepackage[utf8]{inputenc}
\usepackage[T1]{fontenc}
\usepackage{lmodern}

\usepackage{hyperref}
\hypersetup{hypertexnames=false}
\usepackage{xurl}
\usepackage{booktabs}
\usepackage{amsfonts}
\usepackage{amsmath}
\usepackage{amssymb}
\usepackage{mathtools}

\usepackage{graphicx}
\usepackage{tikz}
\usetikzlibrary{arrows.meta,positioning,fit,backgrounds,calc}
\usepackage{microtype}

\usepackage{algorithm}
\usepackage{algpseudocode}

\usepackage{enumitem}
\usepackage{listings}
\usepackage{multirow}
\usepackage{tabularx}

\usepackage[numbers,compress]{natbib}
\usepackage{doi}
\usepackage{placeins}

\newcolumntype{Y}{>{\raggedright\arraybackslash}X}

\definecolor{layerMemory}{HTML}{DCEBFF}
\definecolor{layerCapability}{HTML}{E8F5E9}
\definecolor{layerContext}{HTML}{FFF3D6}
\definecolor{layerExecution}{HTML}{F3E5F5}
\definecolor{lineDark}{HTML}{34495E}
\definecolor{textDark}{HTML}{1F2933}
\definecolor{codeBg}{HTML}{F8FAFC}
\definecolor{codeRule}{HTML}{CBD5E1}
\definecolor{codeKeyword}{HTML}{1D4ED8}
\definecolor{codeComment}{HTML}{64748B}

\lstdefinestyle{runtime}{
	basicstyle=\ttfamily\footnotesize,
	breaklines=true,
	columns=fullflexible,
	keepspaces=true,
	showstringspaces=false,
	frame=single,
	backgroundcolor=\color{codeBg},
	rulecolor=\color{codeRule},
	framerule=0.45pt,
	framesep=6pt,
	xleftmargin=0.75em,
	xrightmargin=0.75em,
	aboveskip=0.8em,
	belowskip=0.8em
}
\tikzset{
	ckb_box/.style={
		draw=lineDark,
		rounded corners=3pt,
		very thick,
		align=center,
		font=\sffamily\small,
		text=textDark,
		minimum height=0.82cm,
		inner xsep=7pt,
		inner ysep=5pt
	},
	ckb_memory/.style={ckb_box, fill=layerMemory, minimum width=4.8cm},
	ckb_capability/.style={ckb_box, fill=layerCapability, minimum width=4.8cm, font=\sffamily\small\bfseries},
	ckb_context/.style={ckb_box, fill=layerContext, minimum width=4.8cm},
	ckb_execution/.style={ckb_box, fill=layerExecution, minimum width=4.8cm},
	ckb_artifact/.style={ckb_box, fill=white, minimum width=4.5cm},
	ckb_note/.style={ckb_box, fill=layerCapability!35, align=left, font=\sffamily\scriptsize},
	ckb_arrow/.style={-{Latex[length=2.5mm]}, very thick, draw=lineDark},
	ckb_light_arrow/.style={-{Latex[length=2.0mm]}, thick, draw=lineDark!75}
}

\title{Knowledge-Centric Information Systems}

\newif\ifuniqueAffiliation
\uniqueAffiliationtrue

\ifuniqueAffiliation
\author{ \href{https://orcid.org/0009-0008-0201-2984}{\includegraphics[scale=0.06]{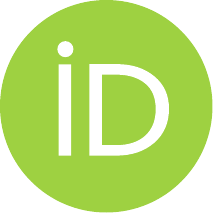}\hspace{1mm}Mariano Garralda-Barrio}\thanks{Independent Researcher / Investigador Independiente.} \\
Independent Researcher\\
Lleida, Spain \\
\texttt{mariano.garralda.r@gmail.com} \\
}
\fi

\renewcommand{\headeright}{}

\renewcommand{\undertitle}{Generalizing Data Engineering Principles for Executable Organizational Knowledge}

\begin{document}
\setlength{\headheight}{22.5pt}
\addtolength{\topmargin}{-8.5pt}
\maketitle

\begin{abstract}
For decades, data engineering has developed mature architectural principles for integrating, governing, validating, cataloging, and serving organizational data. The rise of large language models does not eliminate these concerns; it exposes a broader version of them. Organizational knowledge is becoming executable infrastructure: systems increasingly retrieve it, assemble it, reason over it, and act on it. This paper argues that enterprise artificial intelligence (AI) systems suggest a transition toward an architectural discipline for representing, maintaining, governing, and operationally delivering organizational knowledge. We refer to this discipline as \emph{knowledge architecture}. We offer a conceptual model and taxonomy showing how classical data-engineering guarantees must be redefined when the managed unit shifts from records to knowledge artifacts: extract, transform, and load (ETL) becomes knowledge ingestion, change-data capture (CDC) becomes knowledge change detection, lineage becomes provenance, catalogs become knowledge catalogs, materialized views become knowledge views, and medallion architectures become raw--curated--operational knowledge layers. Emerging formats such as large language model (LLM) Wiki and the Open Knowledge Format (OKF) are treated as early evidence of this transition, not as its endpoint. The central claim is that knowledge architecture becomes useful when organizational knowledge ceases to be a passive information resource and becomes an operational asset used by humans, agents, workflows, and models to execute work.
\end{abstract}


\section{Introduction}

Modern enterprises have spent the last several decades learning how to manage data as an architectural asset. The resulting discipline includes ingestion pipelines, warehouses, lakes, lakehouses, change-data capture, metadata catalogs, governance processes, lineage systems, quality controls, event streams, and consumption interfaces \cite{inmon2005building,kimball2013toolkit,kleppmann2017ddia,armbrust2021lakehouse}. These patterns emerged because organizations had too much data, in too many systems, with too many downstream consumers.

A similar transition is now occurring for organizational knowledge. Large language models (LLMs), retrieval-augmented generation (RAG), semantic search, agentic workflows, and tool-calling systems make it possible to consume information that previously remained outside classical data pipelines \cite{lewis2020rag,yao2023react,park2023generative}. Enterprise knowledge is no longer limited to records in databases. It includes manuals, design documents, images, slide decks, spreadsheets, tickets, contracts, source code, application programming interface (API) specifications, metrics, chat histories, decisions, runbooks, playbooks, and tacit assumptions embedded in workflows.

The architectural problem is changing. The question is no longer only how to process data, but how to represent, maintain, govern, synchronize, validate, and serve heterogeneous knowledge so that humans, agents, workflows, and models can use it reliably. We use \emph{architecture} in the established sense of organizing system structures, stakeholder concerns, viewpoints, and design decisions for complex software-intensive and information systems \cite{zachman1987framework,ieee2000recommended,iso2011architecture,bass2022software}. We call the resulting discipline \emph{knowledge architecture}: an architectural response to the shift from knowledge as a passive resource that humans consult to knowledge as executable infrastructure that systems use to decide, assemble context, call tools, and perform work.

\begin{quote}
\textbf{Definition 1 (Executable organizational knowledge).} Executable organizational knowledge is organizational knowledge that is directly consumable by humans, models, agents, workflows, or applications to influence or perform operational behavior. It is not executable in the narrow sense of program code; it is executable because it can shape decisions, context assembly, tool use, or workflow execution.
\end{quote}

\begin{quote}
\textbf{Definition 2 (Knowledge Architecture).} Knowledge Architecture is the discipline concerned with defining, implementing, and governing the architectural guarantees required for executable organizational knowledge.
\end{quote}

\begin{quote}
\textbf{Proposition 1 (From data-centric to knowledge-centric systems).} When heterogeneous organizational knowledge is consumed operationally by humans, models, agents, workflows, or applications, information systems require architectural guarantees for knowledge artifacts analogous to, but semantically broader than, the guarantees that data architecture provides for records and tables.
\end{quote}

The core thesis is deliberately conservative: knowledge architecture does not replace data architecture. Data architecture remains essential for structured data and analytical systems. Knowledge architecture emerges as a higher-level layer that incorporates data while also covering non-tabular, multimodal, semantic, and operational artifacts. The mature principles of data engineering remain valuable, but they must be lifted to a new unit of management: the \emph{knowledge artifact}. This unit shift distinguishes the proposal from a simple relabeling of knowledge management or enterprise architecture.

Recent developments suggest that this transition is already underway. Karpathy's LLM Wiki pattern proposes a curated markdown knowledge base that language models can read and maintain \cite{karpathy2026llmwiki}. Google Cloud's Open Knowledge Format (OKF) formalizes part of this pattern as a portable, interoperable format for agent-readable knowledge bundles \cite{mcveety2026okf,googlecloud2026okfspec}. These initiatives do not solve the entire knowledge-architecture problem, but they make the broader shift visible: organizations are beginning to treat knowledge representation and interoperability as infrastructure concerns.

This vision paper makes three contributions:

\begin{itemize}[leftmargin=*]
  \item \textbf{A paradigm framing}: we characterize knowledge architecture as the evolution of data architecture for knowledge-centric information systems, where the managed unit shifts from data records to heterogeneous knowledge artifacts.
  \item \textbf{A guarantee taxonomy}: we map established data-engineering guarantees to their knowledge-level reformulations, identifying why their semantics change.
  \item \textbf{An architectural agenda}: we outline open questions for representation, ingestion, synchronization, provenance, governance, validation, observability, retrieval, and interoperability in systems that act on executable knowledge.
\end{itemize}

\section{Motivation: From Passive Knowledge to Executable Infrastructure}

Data engineering emerged because operational systems were not enough. Organizations needed ways to integrate information across applications, produce consistent analytical views, govern access, track transformations, and make data reusable. Data warehouses, data lakes, and lakehouses responded to this need by providing architectural patterns for persistence, transformation, and consumption \cite{inmon2005building,kimball2013toolkit,armbrust2021lakehouse}.

AI systems that operationalize heterogeneous organizational knowledge create a different pressure. Enterprises already possessed large quantities of unstructured and semi-structured information, but most of it was passive. Policies, architecture decisions, onboarding notes, bug discussions, product requirements, incident reports, and customer conversations were documents that humans could consult. They were rarely treated as continuously maintained operational inputs for autonomous or semi-autonomous systems.

Large language models make this latent problem visible because they turn knowledge into a possible execution substrate. They can read documents, synthesize across artifacts, generate summaries, call tools, and operate over mixed textual and structured contexts. As a result, previously passive information becomes a possible input to decisions and actions. The bottleneck moves from model capability to knowledge readiness: whether the right knowledge exists, is current, is governed, is traceable, and can be assembled for the consumer at hand.

Recent work on governed agent-runtime evolution and goal-oriented dialogue runtimes has examined how intelligent systems can move from conversational interaction toward governed operational behavior \cite{garraldabarrio2026governedevolutionagentruntimes,garraldabarrio2026taskguidedconversationalgraphsgoaloriented}. This paper addresses the complementary architectural question: what knowledge infrastructure is required when such systems depend on organizational knowledge as an executable asset?

The historical progression of enterprise architectures has not replaced previous architectural concerns; it has progressively elevated them as the managed asset became richer in semantics. Data engineering emerged when organizations had too much data to manage manually. Knowledge architecture emerges when organizational knowledge becomes executable infrastructure.

The claim is not that knowledge management is new. Organizational memory and knowledge-management research have long studied how institutions create, store, transfer, and apply knowledge \cite{walsh1991organizational,stein1995organizational,nonaka1995knowledge,alavi2001review,davenport1998working}. What is new is the operational pressure created by AI systems that can directly consume and act on broad knowledge surfaces. Knowledge is becoming executable context, and executable context requires architectural guarantees: provenance, freshness, governance, quality, synchronization, observability, and controlled delivery.

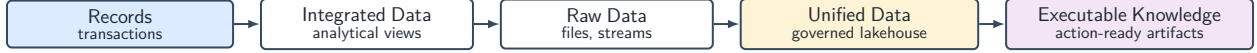
\begin{figure}[t]
\centering
\resizebox{1\linewidth}{!}{%
\begin{tikzpicture}[node distance=0.55cm, every node/.style={font=\sffamily}]
  \node[ckb_memory, text width=2.0cm] (db) {\large Records\\\normalsize transactions};
  \node[ckb_artifact, right=of db, text width=3.45cm] (dw) {\large Integrated Data\\\normalsize analytical views};
  \node[ckb_artifact, right=of dw, text width=2.45cm] (dl) {\large Raw Data\\\normalsize files, streams};
  \node[ckb_context, right=of dl, text width=4.55cm] (lh) {\large Unified Data\\\normalsize governed lakehouse};
  \node[ckb_execution, right=of lh, text width=4.75cm] (ka) {\large Executable Knowledge\\\normalsize action-ready artifacts};
  \draw[ckb_arrow] (db) -- (dw);
  \draw[ckb_arrow] (dw) -- (dl);
  \draw[ckb_arrow] (dl) -- (lh);
  \draw[ckb_arrow] (lh) -- (ka);
\end{tikzpicture}%
}
\caption{A historical progression of managed enterprise assets: architectural concerns are elevated rather than replaced as the asset becomes richer in semantics.}
\label{fig:historical-expansion}
\end{figure}

\section{The Unit Shift: From Record to Knowledge Artifact}

Classical data architecture assumes relatively stable units: records, tables, schemas, streams, files, and events. These units may be complex, but they are usually explicit, typed, and machine-managed. Knowledge-centric systems operate over a broader and less regular unit: the \emph{knowledge artifact}.

A knowledge artifact is any persistent item that contributes to organizational understanding or execution. It may be structured, semi-structured, unstructured, or multimodal. Examples include a database table, a contract clause, an architectural decision record, a monitoring dashboard, a support ticket, a code module, a spreadsheet, a diagram, a presentation, an incident postmortem, a meeting transcript, or an interface specification.

The knowledge artifact is the managed unit because it is the smallest practical object over which architectural responsibilities can be assigned. A record can be typed, validated, versioned, and traced using database or pipeline mechanisms. A knowledge artifact requires analogous controls, but the controls must also preserve meaning: scope, source evidence, authoritativeness, semantic relations, temporal validity, sensitivity, and intended operational use. Without such a unit, knowledge systems collapse into either undifferentiated document collections or transient runtime prompts.

This shift changes the architecture in four ways. First, the boundary of managed content expands beyond databases. Second, the meaning of an artifact often depends on context, provenance, and relationships to other artifacts. Third, consumption is no longer limited to queries and dashboards; it includes retrieval, synthesis, reasoning, tool use, and agentic execution. Fourth, artifacts must be governed for action: when a system uses a runbook, policy, diagram, or decision record to execute work, the artifact becomes part of an operational control surface.

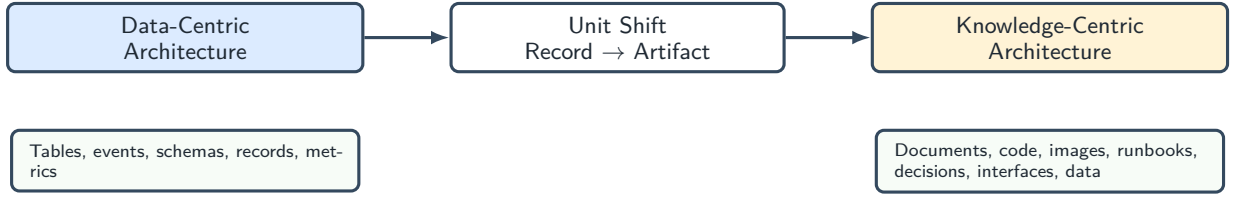
\begin{figure}[t]
\centering
\resizebox{0.98\linewidth}{!}{%
\begin{tikzpicture}[node distance=1.15cm, every node/.style={font=\sffamily}]
  \node[ckb_memory] (data) {Data-Centric\\Architecture};
  \node[ckb_artifact, right=of data] (shift) {Unit Shift\\Record $\rightarrow$ Artifact};
  \node[ckb_context, right=of shift] (knowledge) {Knowledge-Centric\\Architecture};
  \draw[ckb_arrow] (data) -- (shift);
  \draw[ckb_arrow] (shift) -- (knowledge);
  \node[ckb_note, below=0.75cm of data, text width=4.2cm] {Tables, events, schemas, records, metrics};
  \node[ckb_note, below=0.75cm of knowledge, text width=4.2cm] {Documents, code, images, runbooks, decisions, interfaces, data};
\end{tikzpicture}%
}
\caption{The unit shift from structured records to heterogeneous knowledge artifacts expands the scope of architectural responsibility.}
\label{fig:unit-shift}
\end{figure}

\section{Why Data Architecture Alone Is No Longer Sufficient}

The argument is not that data architecture failed. On the contrary, knowledge architecture inherits its most important achievement: the idea that organizational assets require explicit architectural guarantees before they can be reused safely. Data engineering made data reliable by developing guarantees for consistency, lineage, quality, governance, observability, interoperability, and controlled consumption.

Those guarantees do not transfer unchanged when the managed asset becomes executable knowledge. Data architecture is well suited to questions such as whether a record changed, whether a schema is valid, whether a table was transformed, or whether a metric is fresh. Knowledge-centric systems must ask additional questions: whether the meaning of a policy changed, whether a procedure remains authoritative, whether two artifacts contradict each other, whether a synthesized answer preserves its evidential basis, whether a diagram is still valid for a runtime environment, and whether a retrieved artifact is appropriate for an agent about to take action.

The analogy with data engineering is useful but insufficient. Knowledge architecture is not a replacement for data architecture, but a semantic generalization of its architectural guarantees. The same families of concern reappear, but their semantics change because knowledge artifacts carry meaning, authority, context, and operational consequences. The architectural concern survives; the managed asset changes.

\section{A Taxonomy of Knowledge Architecture Guarantees}

The strongest reason to treat knowledge architecture as an evolution of data architecture is that familiar guarantees recur under changed semantics. The problems are recognizable, but they re-emerge over a richer managed asset. The solutions differ because the problem structure is no longer only about data movement, storage, or analytical consumption; it is about ensuring that executable knowledge remains trustworthy when used by humans, models, agents, and workflows.

\begin{table}[t]
\centering
\small
\begin{tabularx}{\linewidth}{p{0.20\linewidth}p{0.24\linewidth}Y}
\toprule
\textbf{Data Guarantee} & \textbf{Knowledge Guarantee} & \textbf{Semantic Shift} \\
\midrule
Ingestion & Knowledge ingestion & Artifacts must be parsed, chunked, linked, summarized, and enriched without losing meaning or evidence. \\
Storage & Knowledge repository & Repositories must preserve raw artifacts, curated knowledge, semantic relations, and operational projections. \\
Change & Knowledge change detection & The relevant change may be semantic: a policy, authority, procedure, dependency, or interpretation may change even when structure does not. \\
Metadata & Knowledge metadata & Metadata must encode ownership, scope, validity, authority, sensitivity, semantic type, and intended operational use. \\
Cataloging & Knowledge catalog & Discovery must include fitness for action, not only discoverability or descriptive metadata. \\
Lineage & Knowledge provenance & Provenance must trace sources, extraction, synthesis, human approval, and runtime assembly into context. \\
Quality & Knowledge quality & Quality includes freshness, consistency, grounding, contradiction management, completeness, and operational safety. \\
Governance & Knowledge governance & Governance must control who can author, approve, activate, retrieve, or expose knowledge to agents and workflows. \\
Materialization & Knowledge views & Views may be task-specific context bundles, decision-support packs, or agent-readable projections. \\
CQRS & Knowledge read/write split & Authoring, curation, approval, retrieval, and runtime consumption require different controls, latencies, and consistency expectations \cite{fowler2011cqrs}. \\
Layering & Knowledge maturity layers & Movement between layers includes extraction, semantic linking, validation, synthesis, approval, and readiness for execution. \\
\bottomrule
\end{tabularx}
\caption{Data-engineering guarantees recur in knowledge architecture with changed semantics.}
\label{tab:taxonomy}
\end{table}

This taxonomy is intentionally architectural rather than product-specific. It does not claim that every data-engineering pattern transfers directly. Instead, it shows where the guarantee must be redefined. Change detection is no longer only row-level change-data capture; it includes semantic drift, authority changes, and procedural obsolescence. Lineage is no longer only table-to-table transformation; it includes extraction, model-mediated synthesis, human approval, and context assembly. Materialization and command query responsibility segregation (CQRS) are related but distinct: materialization concerns consumer-ready projections, while CQRS separates the model used to update a system from the model used to read from it \cite{fowler2011cqrs}. At the knowledge level, this means separating authoring and curation flows from retrieval and runtime consumption. Quality is no longer only validity or completeness; it includes grounding, contradiction management, and safety for action.

\section{Emerging Evidence: LLM Wiki and Open Knowledge Format}

LLM Wiki and OKF are useful examples because they point toward the representation and interoperability layer of knowledge architecture. The LLM Wiki pattern treats markdown-based knowledge bases as maintainable resources that agents can read, update, and cross-reference \cite{karpathy2026llmwiki}. OKF formalizes part of this direction as a portable bundle format for organizational knowledge, using markdown files and structured metadata so that tools and agents can exchange knowledge more easily \cite{mcveety2026okf,googlecloud2026okfspec}.

The architectural lesson is not that OKF is a complete knowledge platform. A format can help standardize representation and interchange, but it does not by itself solve semantic synchronization, provenance across synthesis, governance of executable knowledge, quality, observability, lifecycle management, or operational assembly. This is a boundary clarification rather than a criticism: representation is one layer of the architecture, not the whole architecture.

This distinction is central to the argument of this paper. If OKF or any successor standard disappeared, the underlying problem would remain: organizations still need ways to turn heterogeneous artifacts into reliable, governed, interoperable, and operationally useful knowledge. Standards are symptoms and accelerators of the paradigm shift, not the shift itself.

\section{Industrial Signals of Convergence}

The argument above is conceptual, but it is not detached from practice. Several contemporary systems suggest that industry is converging on the same architectural pressure: knowledge must be made available to systems that plan, generate, retrieve, and act.

First, GitHub Copilot Workspace treats repository issues, code, and project context as inputs to an artificial-intelligence-assisted development workflow that can move from task description to proposed implementation artifacts \cite{githubnext2024workspace}. Second, Microsoft Copilot Studio exposes enterprise knowledge sources such as files, SharePoint, Dataverse, Dynamics 365, Salesforce, ServiceNow, Azure AI Search, and Structured Query Language (SQL)-backed sources to agents, making knowledge selection and grounding an explicit agent-design concern \cite{microsoft2026copilotknowledge}. Third, Google Cloud's OKF and Knowledge Catalog direction makes knowledge bundles ingestible and servable to agents, pointing toward interoperability and representation as infrastructure concerns \cite{mcveety2026okf,googlecloud2026okfspec}.

These examples should not be read as evidence that knowledge architecture is already a settled discipline. They are better understood as early industrial signals: different vendors are independently exposing the same underlying requirement to represent, govern, retrieve, and operationalize heterogeneous knowledge for executable systems.

\section{Reference Model}

A knowledge architecture can be described as a layered system that connects raw organizational artifacts to operational consumers.

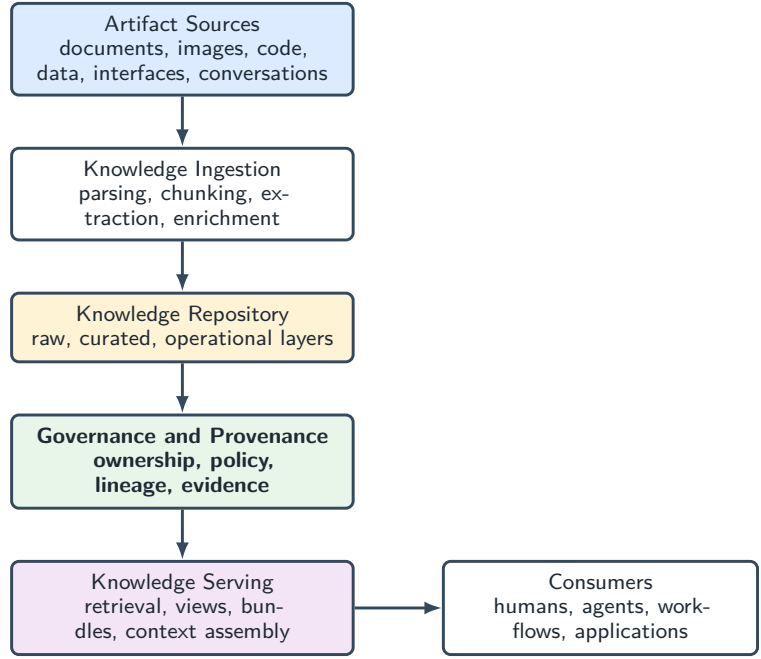
\begin{figure}[t]
\centering
\resizebox{0.6\linewidth}{!}{%
\begin{tikzpicture}[node distance=0.72cm, every node/.style={font=\sffamily}]
  \node[ckb_memory, text width=4.4cm] (sources) {Artifact Sources\\\footnotesize documents, images, code, data, interfaces, conversations};
  \node[ckb_artifact, below=of sources, text width=4.4cm] (ingest) {Knowledge Ingestion\\\footnotesize parsing, chunking, extraction, enrichment};
  \node[ckb_context, below=of ingest, text width=4.4cm] (repo) {Knowledge Repository\\\footnotesize raw, curated, operational layers};
  \node[ckb_capability, below=of repo, text width=4.4cm] (govern) {Governance and Provenance\\\footnotesize ownership, policy, lineage, evidence};
  \node[ckb_execution, below=of govern, text width=4.4cm] (serve) {Knowledge Serving\\\footnotesize retrieval, views, bundles, context assembly};
  \node[ckb_artifact, right=1.25cm of serve, text width=4.0cm] (consumers) {Consumers\\\footnotesize humans, agents, workflows, applications};
  \draw[ckb_arrow] (sources) -- (ingest);
  \draw[ckb_arrow] (ingest) -- (repo);
  \draw[ckb_arrow] (repo) -- (govern);
  \draw[ckb_arrow] (govern) -- (serve);
  \draw[ckb_arrow] (serve) -- (consumers);
\end{tikzpicture}%
}
\caption{Reference model for knowledge architecture: heterogeneous artifacts are ingested, governed, served, and consumed operationally.}
\label{fig:reference-model}
\end{figure}

The model separates five concerns. \emph{Sources} are the systems and artifacts from which knowledge is derived. \emph{Ingestion} converts artifacts into machine-usable representations through parsing, extraction, annotation, embedding, linking, or summarization. \emph{Repositories} persist knowledge at different maturity levels. \emph{Governance and provenance} establish trust, ownership, access, lineage, and accountability. \emph{Serving} exposes knowledge through retrieval interfaces, bundles, views, APIs, and runtime context assembly.

\section{Key Architectural Concerns}

\subsection{Knowledge Representation}

Knowledge representation concerns the forms in which knowledge artifacts and their relations are encoded. Markdown bundles, schemas, ontologies, knowledge graphs, embeddings, vector indexes, relational tables, and object stores may all participate. The architectural question is not which representation wins, but how multiple representations coexist and remain interoperable.

\subsection{Knowledge Ingestion}

Knowledge ingestion generalizes extract, transform, and load. It includes document parsing, optical character recognition (OCR), table extraction, multimodal processing, code analysis, metadata enrichment, entity linking, embedding generation, and transformation into reusable knowledge units. Unlike classical extract, transform, and load, ingestion may produce both symbolic and statistical representations.

\subsection{Knowledge Synchronization}

Knowledge artifacts change continuously. Policies are updated, interfaces evolve, spreadsheets are replaced, code is refactored, and decisions become obsolete. Knowledge change detection generalizes change-data capture beyond databases. It must detect changes in meaning, not merely changes in bytes.

\subsection{Knowledge Provenance}

Knowledge provenance generalizes data lineage. A knowledge claim should be traceable to source artifacts, extraction steps, transformations, validations, approvals, and consumers. Provenance becomes especially important when large language models synthesize new artifacts from existing ones, because synthesized knowledge can obscure its evidential basis.

\subsection{Knowledge Quality}

Knowledge quality includes accuracy, freshness, consistency, completeness, grounding, non-duplication, scope validity, and operational suitability. Quality checks may combine deterministic rules, human review, automated consistency tests, retrieval evaluation, and model-based validation. The target is not perfect truth, but controlled fitness for use.

\subsection{Knowledge Governance}

Knowledge governance defines who can create, modify, approve, access, publish, archive, or delete knowledge artifacts. It must handle sensitivity, intellectual property, regulatory constraints, privacy, jurisdiction, and accountability. In agentic systems, governance also determines which knowledge can be activated for which task.

\subsection{Knowledge Serving}

Knowledge serving concerns how curated knowledge reaches consumers. Serving may use search, semantic retrieval, graph traversal, generated summaries, interface endpoints, context windows, Open-Knowledge-Format-like bundles, or task-specific knowledge views. The serving layer should separate persistent knowledge from ephemeral runtime context.

\section{Medallion Knowledge Architecture}

One concrete example of pattern elevation is the medallion architecture. In data systems, medallion architectures distinguish raw, cleaned, and curated layers, often called bronze, silver, and gold \cite{armbrust2021lakehouse}. Knowledge systems can adopt a similar structure while changing the semantics of each layer.

\begin{table}[t]
\centering
\small
\begin{tabularx}{\linewidth}{p{0.2\linewidth}p{0.34\linewidth}Y}
\toprule
\textbf{Layer} & \textbf{Knowledge Role} & \textbf{Examples} \\
\midrule
Raw knowledge & Captured artifacts with minimal transformation & documents, slides, chat logs, tickets, source files, spreadsheets, tables \\
Curated knowledge & Normalized, enriched, linked, and reviewed artifacts & Extracted entities, summaries, metadata, provenance links, validated snippets \\
Operational knowledge & Consumer-ready knowledge views and bundles & Agent context packs, runbook views, decision-support bundles, compliance knowledge sets \\
\bottomrule
\end{tabularx}
\caption{Medallion-style layering for knowledge systems: raw artifacts become curated and operational knowledge.}
\label{tab:medallion}
\end{table}

This example illustrates the broader thesis. The data pattern remains recognizable, but the managed object changes. Transformation is no longer only a schema operation; it may include summarization, semantic linking, policy annotation, and provenance preservation.

\section{Boundary of the Claim}

This paper does not argue that knowledge architecture is entirely new. Knowledge management, organizational memory, semantic web research, enterprise architecture, information retrieval, software architecture, and data engineering all contribute foundations \cite{zachman1987framework,boehm1988spiral,garlan1993introduction,ieee2000recommended,iso2011architecture,bass2022software,walsh1991organizational,nonaka1995knowledge,bernerslee2001semantic,opengroup2018togaf}. The claim is narrower: AI systems that operationalize heterogeneous organizational knowledge are forcing these traditions into a shared operational architecture, where heterogeneous knowledge must be maintained and served with the reliability expected from modern data platforms.

Ontologies, knowledge graphs, and Semantic Web standards provide important foundations for representing machine-readable meaning and relationships. The argument here is not that such representations are insufficient, but that executable organizational knowledge also requires lifecycle guarantees for ingestion, synchronization, provenance, governance, quality, observability, and operational serving.

Knowledge architecture is differentiated not by the novelty of every component, but by the conjunction of three conditions: the managed asset is heterogeneous knowledge artifacts; the consumer includes models, agents, and workflows; and the consumption mode includes operational execution. Knowledge management studied organizational knowledge as a managerial and socio-technical asset. Semantic Web research studied machine-readable meaning and relationships. Enterprise architecture studied organizational structure and capabilities. Data engineering studied reliable pipelines and platforms. Knowledge architecture sits at their intersection when executable systems require governed, traceable, interoperable knowledge as infrastructure.

The term \emph{knowledge architecture} is used deliberately. Alternatives such as context architecture or artificial intelligence knowledge infrastructure name important downstream or implementation-specific concerns. Context architecture emphasizes what is assembled for a particular execution episode; AI knowledge infrastructure emphasizes the technical substrate used by AI systems. Knowledge architecture names the broader architectural object: the organizational knowledge asset itself, together with the guarantees that make it usable across changing humans, models, agents, workflows, and applications.

Nor does this paper claim that large language models create organizational knowledge. They make knowledge easier to consume, transform, and activate, but the architectural responsibility remains organizational. Knowledge must still be sourced, governed, validated, traced, and retired.

Finally, knowledge architecture should not be reduced to retrieval-augmented generation. RAG is a serving technique, not a complete architecture \cite{lewis2020rag}. Reliable knowledge systems require upstream concerns before retrieval and downstream concerns after generation.

\section{Research Agenda}

The proposed framing raises several research questions:

\begin{itemize}[leftmargin=*]
  \item \textbf{Representation}: what minimal metadata and relationship models are required for interoperable knowledge artifacts?
  \item \textbf{Synchronization}: how can knowledge systems detect semantic drift across documents, code, interfaces, and databases?
  \item \textbf{Provenance}: how should synthesized knowledge preserve evidence, transformations, and responsibility?
  \item \textbf{Quality}: what metrics determine whether a knowledge artifact is fit for operational use?
  \item \textbf{Governance}: how should access, approval, retention, and policy controls apply to knowledge consumed by agents?
  \item \textbf{Evaluation}: how can organizations measure the reliability, freshness, coverage, and usefulness of knowledge architectures?
\end{itemize}

\section{Discussion}

The most important implication is that enterprise AI should not be designed only around models or agents. Models consume knowledge, but they do not define the institutional knowledge boundary. Agents execute tasks, but they should not become the sole containers of organizational memory. Durable knowledge must be represented independently from the transient systems that consume it.

This perspective also reduces technological overfitting. OKF, knowledge graphs, vector databases, Model Context Protocol servers, agent memories, and semantic indexes may all be useful, but each solves only part of the architectural space. A mature knowledge architecture will likely combine several of them, just as mature data platforms combine databases, streams, warehouses, lakes, catalogs, governance tools, and serving layers.

The practical recommendation is to reuse the lessons of data engineering without assuming equivalence. Knowledge systems need ingestion, catalogs, provenance, quality, governance, synchronization, and serving, but each guarantee must be reinterpreted for meaning-bearing artifacts that may be synthesized, contradicted, approved, assembled into context, and used for action. The goal is not to move data more efficiently; it is to make executable organizational knowledge trustworthy enough for operational use.

\section{Conclusion}

This paper proposed knowledge architecture as the semantic generalization of data-architecture guarantees for knowledge-centric information systems: systems that operationalize heterogeneous organizational knowledge. The central observation is that the classical guarantees of data engineering do not disappear when organizations build such systems; they must be redefined at a higher level of abstraction. The managed unit shifts from records and tables to heterogeneous knowledge artifacts, while ingestion, storage, synchronization, quality, governance, provenance, observability, interoperability, and consumption remain under changed semantics.

Initiatives such as LLM Wiki and OKF provide early evidence that the industry is beginning to standardize the representation and interchange of organizational knowledge. Yet representation is only one layer. The broader challenge is architectural: building systems that can maintain, govern, validate, and serve knowledge reliably across changing models, agents, tools, and workflows.

The claim is simple: data remains essential, but knowledge is becoming the next managed architectural asset. If data engineering made organizational data reusable, knowledge architecture aims to make organizational knowledge trustworthy enough to execute.

\FloatBarrier

\section*{Acknowledgements}
The author acknowledges the Laboratorio de Innovación Aplicada (L2IA) at Minsait (Indra Group) for fostering an environment that encourages scientific exploration in AI systems, software architecture, and organizational knowledge for intelligent systems.

\bibliographystyle{unsrtnat}
\bibliography{reference}

\end{document}